\documentclass[prl,twocolumn,fleqn]{revtex4}
\usepackage{amscd,verbatim}
\usepackage[all]{xy}
\usepackage{graphicx}
\usepackage{amsmath}
\usepackage[psamsfonts]{amssymb}
\usepackage{euscript}

\usepackage{latexsym}

\def\hsp{,\hspace{.7cm}}

\def\th#1#2{\ensuremath{\theta_{#1#2}}}

\def\cp#1#2#3{\hbox{\rm cos}^#1(\th#2#3)}
\def\sp#1#2#3{\hbox{\rm sin}^#1(\th#2#3)}

\def\spp#1#2#3#4{\hbox{\rm sin}^#1(#2\th#3#4)}

\def\m#1#2{\ensuremath{\Delta M_{#1#2}^2}}
\def\mn#1#2{\ensuremath{|\Delta M_{#1#2}^2}|}

\def\meff{\ensuremath{\Delta M^2_{\rm{eff}}}}
\renewcommand{\cos}{\textrm{cos}}
\renewcommand{\sin}{\textrm{sin}}

\renewcommand{\(}{\begin{equation}}
\renewcommand{\)}{end{equation} \vspace{-.05in}\linebreak}

\newcounter{saveeqn}
\newcounter{savealpheqn}

\newcommand{\alpheqn}{\setcounter{saveeqn}{\value{equation}}%
  \stepcounter{saveeqn}\setcounter{equation}{0}%
  \renewcommand{\theequation}{\mbox{\arabic{section}.\arabic{saveeqn}
\alph{equation}}}
  \renewcommand{\)}{\end{equation}}}
\def\part#1{\frac{\partial}{\partial{#1}}}%
\def\group#1{\refstepcounter{equation}\setcounter{saveeqn}
 {\value{equation}}%
  \label{#1}\setcounter{equation}{0}%
\renewcommand{\theequation}{\mbox{\arabic{section}.\arabic{saveeqn}
\alph{equation}}}
  \renewcommand{\)}{\end{equation}}}
\newcommand{\reseteqn}{\setcounter{equation}{\value{saveeqn}}%
  \renewcommand{\theequation}{\arabic{section}.\arabic{equation}}%
  \renewcommand{\)}{\end{equation}}}

\newcommand{\aalpheqn}{\setcounter{saveeqn}{\value{equation}}%
  \stepcounter{saveeqn}\setcounter{equation}{0}%
  \renewcommand{\theequation}{\mbox{
        \Alph{subsection}.\arabic{saveeqn}\alph{equation}}}
   \renewcommand{\)}{\end{equation}}}
\newcommand{\areseteqn}{\setcounter{equation}{\value{saveeqn}}%
  \renewcommand{\theequation}{\Alph{subsection}.\arabic{equation}}%
  \renewcommand{\)}{\end{equation}}}

\renewcommand{\thefootnote}{\alph{footnote}}
\renewcommand{\(}{\begin{equation}}
\renewcommand{\)}{\end{equation}}
\newcommand{\ba}{\begin{eqnarray}}
\newcommand{\ea}{\end{eqnarray}}

\newcommand{\bp}{\mathop{\vtop{\ialign{##\crcr
   $\hfil\displaystyle{}\hfil$\crcr\noalign{\kern-13pt\nointerlineskip}
   \BIG{(}\hskip0pt\crcr\noalign{\kern3pt}}}}}
\newcommand{\cbp}{\mathop{\vtop{\ialign{##\crcr
   $\hfil\displaystyle{}\hfil$\crcr\noalign{\kern-13pt\nointerlineskip}
   \BIG{)}\hskip0pt\crcr\noalign{\kern3pt}}}}}
\newcommand{\pa}{\mathop{\vtop{\ialign{##\crcr
    
$\hfil\displaystyle{\oplus}\hfil$\crcr\noalign{\kern+1pt\nointerlineskip 
}
   \hspace{.08in}$^{\alpha=0}$\hskip6pt\crcr\noalign{\kern3pt}}}}}
\renewcommand{\hsp}{,\hspace{.3in}}

\newcommand{\beq}{\begin{equation}}
\newcommand{\eeq}{\end{equation}}

\numberwithin{equation}{section}
\renewcommand{\theequation}{\mbox{\arabic{equation}}}

\def\hsp#1{\hspace{#1in}}

\catcode`\@=11
\def\vereq#1#2{\lower3pt\vbox{\baselineskip1.5pt \lineskip1.5pt
\ialign{$\m@th#1\hfill##\hfil$\crcr#2\crcr\sim\crcr}}}
\catcode`\@=12

\makeatletter
\newcommand\figcaption{\def\@captype{figure}\caption}
\newcommand\tabcaption{\def\@captype{table}\caption}
\makeatother
\renewcommand{\(}{\begin{equation}}
\renewcommand{\)}{\end{equation}}

\def\th#1#2{\ensuremath{\theta_{#1#2}}}

\def\cp#1#2#3{\hbox{\rm cos}^#1(\th#2#3)}
\def\sp#1#2#3{\hbox{\rm sin}^#1(\th#2#3)}

\def\spp#1#2#3#4{\hbox{\rm sin}^#1(#2\th#3#4)}

\def\m#1#2{\ensuremath{\Delta M_{#1#2}^2}}
\def\mn#1#2{\ensuremath{|\Delta M_{#1#2}^2}|}

\def\meff{\ensuremath{\Delta M^2_{\rm{eff}}}}

\renewcommand{\beq}{\begin{equation}}
\renewcommand{\eeq}{\end{equation}}
\newcommand{\bea}{\begin{eqnarray}}
\newcommand{\eea}{\end{eqnarray}}
\newcommand{\beas}{\begin{eqnarray*}}
\newcommand{\eeas}{\end{eqnarray*}}

\newcommand{\bquo}{\begin{quote}}
\newcommand{\enqu}{\end{quote}}

\def\hsp{,\hspace{.2cm}}

\newcommand{\by}{{\mathbf{y}}}
\newcommand{\bx}{{\mathbf{x}}}
\newcommand{\byn}{{\mathbf{y}}^{\mathrm{N}}}
\newcommand{\byi}{{\mathbf{y}}^{\mathrm{I}}}

\newcommand{\pic}{\hspace{-.05cm},\hspace{-.05cm}}

\begin{document}
\def\thefootnote{\fnsymbol{footnote}}

\title{Sensitivity to the Neutrino Mass Hierarchy}

\author{Emilio Ciuffoli, Jarah Evslin and Xinmin Zhang}
\affiliation{Theoretical Physics Division, IHEP, 
CAS, YuQuanLu 19B, Beijing 100049, China 
}

\begin{abstract}
\noindent
In the next decade, a number of experiments will attempt to determine the neutrino mass hierarchy.  Feasibility studies for such experiments generally determine the statistic $\overline{\Delta\chi^2}$.  As the hierarchy is a discrete choice, $\Delta\chi^2$ does not obey a one degree of freedom $\chi^2$ distribution and so the number of $\sigma$'s of confidence of the hierarchy determination is not the square root of $\overline{\Delta\chi^2}$.  We present a simple Bayesian formula for the sensitivity to the hierarchy that can be expected from the median experiment as a function of $\overline{\Delta\chi^2}$.  


\end{abstract}

%
\setcounter{footnote}{0}
\renewcommand{\thefootnote}{\arabic{footnote}}


\maketitle


In the next two decades a number of reactor, accelerator and atmospheric neutrino experiments will attempt to determine the neutrino mass hierarchy, which is the sign of the mass difference $\m31=M^2_3-M^2_1$ where
$M_i$ is the $i$th eigenvalue of the neutrino mass matrix.  
If the sign is positive (negative), one says that the hierarchy is normal (inverted).   Most of these experiments are still in the planning stages, where the key role is played by studies of the sensitivity of a given design to the hierarchy.  

Such studies determine, either analytically or via Monte Carlo simulations, 
\beq
\Delta\chi^2=\chi^2_I-\chi^2_N \label{deltanoi}
\eeq
where $\chi^2_N$ ($\chi^2_I$) is the $\chi^2$ statistic equal to a weighted sum of the squares of the differences between the data and predictions given the normal (inverted) hierarchy, choosing all of the nuisance parameters so as to minimize $\chi^2_N$ ($\chi^2_I$).   The goal of these experiments is not to determine whether each of the hierarchies is consistent with the data, as would be usual in a frequentist approach.  Rather, as it is already well accepted that precisely one of the hierarchies is manifested in nature, the goal of these experiments is to determine {\it{which}} of the hierarchies provides the best fit to the data.  In this paper we will use the test statistic $\Delta\chi^2$ to answer this question as follows.  We will define the best fit hierarchy to be that which yields the lowest value of $\chi^2$, and so the hierarchy determined by the experiment simply corresponds to the sign of $\Delta\chi^2$.

The critical question is then, given $\Delta\chi^2$, what is the sensitivity of a typical experiment to the hierarchy?  In Ref.~\cite{xinoct} the authors showed that the most naive answer, the $p$ value that would be obtained if $\Delta\chi^2$ satisfied a one degree of freedom $\chi^2$ distribution, gives the incorrect answer.  Indeed $\Delta\chi^2$ is not necessarily positive and so such a prescription would not even always be defined.   In this note we will provide an analytic answer (\ref{s}) to this question and will compare our answer to the results of simulations of Daya Bay II and disappearance data at NO$\nu$A.


\vspace{.4cm}

\noindent
{\bf{Nested hypotheses}}

To begin, we will describe just why the $p$ value is not the answer to the question stated above.  Consider $N$ data points $\{y_i\}$ generated by an experiment trying to determine an unknown quantity $x$.  We will use the approximation in which these data points $y_i$ follow a Gaussian distribution peaked at $y_i^{(0)}(x)$ with variance $\sigma_i^2(x)$.   Both $y_i^{(0)}(x)$ and $\sigma_i(x)$ are known functions of $x$.  An experimenter is interested in two hypotheses.  Hypothesis (A) states that $x$ is a real number.  Hypothesis (B) states that $x=x_0$, for a particular real number $x_0$.  Clearly hypothesis (B) is a special case of hypothesis (A), so these hypotheses are said to be nested.  In particular, (B) is obtained from (A) by fixing one, otherwise unconstrained, real number, the number $x$.   

For any given value of $x$, the experimenter can define a statistic $\chi^2(x)$ by simulating the experiment with that value of $x$ and calculating the weighted sum of the squares of differences between his measured and simulated results
\beq
\chi^2(x)=\sum_i\frac{(y_i-y_i^{(0)}(x))^2}{\sigma_i(x)^2}.
\eeq
The experimenter then determines a best fit $\overline{x}$, for which $\chi^2(x)$ is minimized.  He then asks how compatible his results are with the hypothesis (B).  To determine this, he calculates
\beq
\delta\chi^2=\chi^2(x_0)-\chi^2(\overline{x}).
\eeq  
Unlike $\Delta\chi^2$ defined in Eq.~(\ref{deltanoi}), $\delta\chi^2$ is manifestly nonnegative, because $\overline{x}$ is defined so as to give the lowest value of $\chi^2$.    

Just what value of $\delta\chi^2$ should the experimenter expect?  75 years ago Wilks proved \cite{wilks} that if hypothesis (B) is true then $\delta\chi^2$ will obey a $\chi^2$ distribution with a single degree of freedom.    The experimenter can then determine a conditional probability that given (B), the experiment would have gone as badly as it did
\beq
p_W(\delta\chi^2)=\frac{1}{2}\left(1-{\rm{erf}}\left(\sqrt{\frac{\delta\chi^2}{2}}\right)\right).
\eeq
For example, if he found $\delta\chi^2=9$, then $p_W$ would only be about $0.13\%$, and so a frequentist experimenter might conclude that he has ruled out (B) with 3$\sigma$ of confidence.  


\vspace{.4cm}

\noindent
{\bf{Non-nested Hypotheses}}

As described in Ref.~\cite{xinoct}, the determination of the hierarchy is qualitatively different.  The two hypotheses are the normal hierarchy (NH) and the inverted hierarchy (IH).  These hypotheses are not nested, and they correspond to a discrete choice, not the fixing of a degree of freedom.  So the conditions for Wilks' theorem are strongly violated.   As was observed in general in Ref.~\cite{cox} and in this context in Ref.~\cite{xinoct}, this means that the statistic $\Delta\chi^2$ defined in Eq.~(\ref{deltanoi}) does not follow a $\chi^2$ distribution.

Just what distribution does $\Delta\chi^2$ follow?  Let us begin with the simple case in which there are no nuisance parameters, which was applied to a toy model of the hierarchy determination in Ref.~\cite{xinoct}. 

An experiment will produce a set of numbers $\{y_i\}$, which we assemble into a vector $\by$.  The normal and inverted hierarchies yield two theoretical estimates of this vector which we will denote $\byn$ and $\byi$ respectively.  Again let us assume that the measured numbers $y_i$ are normally distributed about their mean with a variance $\sigma_i^2$, which for simplicity we take to be independent of the hierarchy.  Without loss of generality, let us assume for the moment that the true hierarchy is normal.  Then the measured numbers will be
\beq
y_i=y_i^{\mathrm{N}}+\sigma_i g_i
\eeq
where $g_i$ is a standard Gaussian random variable.  

The statistic $\Delta\chi^2$ is then easily determined to be 
\bea
\Delta\chi^2&=&\chi^2_I-\chi^2_N \label{xincont}\\
&=&\sum_i\frac{(y_i-y_i^{\mathrm{I}})^2}{\sigma_i^2}-\sum_i\frac{(y_i-y_i^{\mathrm{N}})^2}{\sigma_i^2}\nonumber\\
&=&\sum_i\frac{(y_i^{\mathrm{N}}+\sigma_i g_i-y_i^{\mathrm{I}})^2-(y_i^{\mathrm{N}}+\sigma_i g_i-y_i^{\mathrm{N}})^2}{\sigma_i^2}\nonumber\\
&=&\sum_i\frac{(y_i^{\mathrm{N}}-y_i^{\mathrm{I}})^2}{\sigma_i^2}+\sum_i\frac{2(y_i^{\mathrm{N}}-y_i^{\mathrm{I}})}{\sigma_i}g_i.\nonumber
\eea 
This identifies $\Delta\chi^2$ as a Gaussian distributed random variable with mean given by the first term on the right hand side
\beq
\overline{\Delta\chi^2}=\sum_i\frac{(y_i^{\mathrm{N}}-y_i^{\mathrm{I}})^2}{\sigma_i^2}
\eeq
and standard deviation given by the second term~\cite{xinoct}
\beq
\sigma_{\Delta\chi^2}=\sqrt{\sum_i\frac{4(y_i^{\mathrm{N}}-y_i^{\mathrm{I}})^2}{\sigma_i^2}}=2\sqrt{\overline{\Delta\chi^2}}. \label{sigeq}
\eeq

Note that $\overline{\Delta\chi^2}$ is the $\Delta\chi^2$ statistic without statistical fluctuations, for example it may be given by the theoretical spectra of $\overline{\nu}_e$ observed at a reactor experiment, of $\nu_\mu$ and $\overline{\nu}_\mu$ at an iron calorimeter atmospheric neutrino experiment, or of $\nu_e$ ($\overline{\nu}_e$) appearance at an accelerator experiment running in the neutrino (antineutrino) mode.  In an atmospheric neutrino experiment one may use the spectra as a function of energy, zenith angle and even the inelasticity of the events \cite{smirnov2013}. $\overline{\Delta\chi^2}$ is the statistic most often reported in the literature.   We will now use Eq. (\ref{sigeq}) to relate  $\overline{\Delta\chi^2}$ to three quantities of interest.

\vspace{.2cm}

\noindent
{\it{What is the probability that the hierarchy which yields the lowest $\chi^2$ is indeed the true hierarchy?}}


Let us first consider the case in which the normal hierarchy is manifested in nature.   The correct hierarchy will be determined by the experiment if $\Delta\chi^2>0$.  The statistic $\Delta\chi^2$ is centered on the positive value $\overline{\Delta\chi^2}$ with a standard deviation of $2\sqrt{\overline{\Delta\chi^2}}$ and so the closest negative value is $\sqrt{\overline{\Delta\chi^2}}/2$ $\sigma$'s from the mean, on one side of the distribution.  For example, if $\overline{\Delta\chi^2}=9$ then a negative value will be excluded at 1.5$\sigma$'s on one side, yielding a probability of successfully determining the hierarchy of 93.3\%, considerably less than the 99.7\% that one may naively suspect just by taking the square root of $\overline{\Delta\chi^2}$.  More generally, the probability of correctly determining the hierarchy is
\beq
p_c(\overline{\Delta\chi^2})=\frac{1}{2}\left(1+{\rm{erf}}\left(\sqrt{\frac{\overline{\Delta\chi^2}}{8}}\right)\right). \label{suc}
\eeq
In a more standard terminology, $p_c$ is the sensitivity to the hierarchy of the binary classification test whose classification function is the sign of $\Delta\chi^2$.  We will refer to it simply as the ``probability of success" in what follows.  In Ref. \cite{hagiwara} the authors obtained a similar result which differs as a result of their formula (5.11) for the probability of success for a given $\Delta\chi^2$.

If instead the inverted hierarchy is correct, the calculation proceeds identically.  As we have approximated $\sigma_i$ to be hierarchy independent, the probability of success is identical for both hierarchies.    This is the quantity quoted in a number of studies such as Refs.~\cite{caojun2,xinag,noisim}.


\vspace{.2cm}

\noindent
{\it{Second, what is the sensitivity of a  typical experiment to the hierarchy?}}

A ``typical experiment" is one in which $|\Delta\chi^2|$ obtains its average value $|\overline{\Delta\chi^2}|$.  As the probability of successfully determining the hierarchy is a monotonic function of $\Delta\chi^2$, the average value of $\Delta\chi^2$ corresponds to the median value of the probability of success and so we will refer to such experiments as median experiments.   The sensitivity of the sign$(\Delta\chi^2)$ test to the hierarchy is the probability that a fit to the correct hierarchy yields a lower value of $\chi^2$ than one to the wrong hierarchy.   Since $|\Delta\chi^2|$ is fixed, this is simply the probability that $\Delta\chi^2$ has the correct sign.  

Again the calculation will proceed identically for both hierarchies, so we may restrict our attention to the case in which the normal hierarchy is correct.  Therefore the question is, given that $\overline{\Delta\chi^2}$ is positive and $\Delta\chi^2$ is equal to either $\overline{\Delta\chi^2}$ or $-\overline{\Delta\chi^2}$, what is the probability $p_v$ that $\Delta\chi^2=\overline{\Delta\chi^2}$.  

Let $L_\pm$ be the likelihood, given the normal hierarchy, that $\Delta\chi^2=\pm\overline{\Delta\chi^2}$, which is easily found using the fact that $\Delta\chi^2$ obeys a normal distribution centered at $\overline{\Delta\chi^2}$ with standard deviation $2\sqrt{\Delta\chi^2}$. Using the fact that the distribution of $\Delta\chi^2$ is odd with respect to a change in the hierarchy, the Bayes factor for the normal hierarchy is
\beq
\frac{L_+}{L_-}=e^{\overline{\Delta\chi^2}/2}.
\eeq
In particular, symmetric Bayesian priors assigning a 50\% chance to each hierarchy yield a probability of success of
\beq
p_v=\frac{L_+}{L_++L_-}=\frac{1}{1+e^{-\overline{\Delta\chi^2}/2}} \label{princ}
\eeq
for median experiments, those in which $|\Delta\chi^2|=|\overline{\Delta\chi^2}|$.  For example, if $\overline{\Delta\chi^2}=9$ then the probability that a median experiment correctly determines the hierarchy will be 98.9\%.  While this is better than the mean probability of success 93.3\%, it still falls noticeably short of the 99.7\% which one might expect from Wilks' theorem.  In Ref.~\cite{blennow} it was noted that the sensitivity (\ref{princ}) is equal to the posterior probability of determining the correct hierarchy.


Given $\overline{\Delta\chi^2}$ determined either from Monte Carlo simulations or from Asimov data, one may express the sensitivity to the hierarchy expected at a median experiment in terms of a number $s$ of standard deviations $\sigma$.  We will convert probabilities into standard deviations using the one-sided Gaussian distribution  
\beq
p_v(\overline{\Delta\chi^2})=\frac{1}{2}\left(1+{\rm{erf}}\left(\frac{s}{\sqrt{2}}\right)\right). \label{peq}
\eeq
While the double-sided Gaussian is also often used in the literature, we have checked that this choice of convention has a small effect on our results.

\begin{figure} 
\begin{center}
\includegraphics[width=2.8in,height=1.3in]{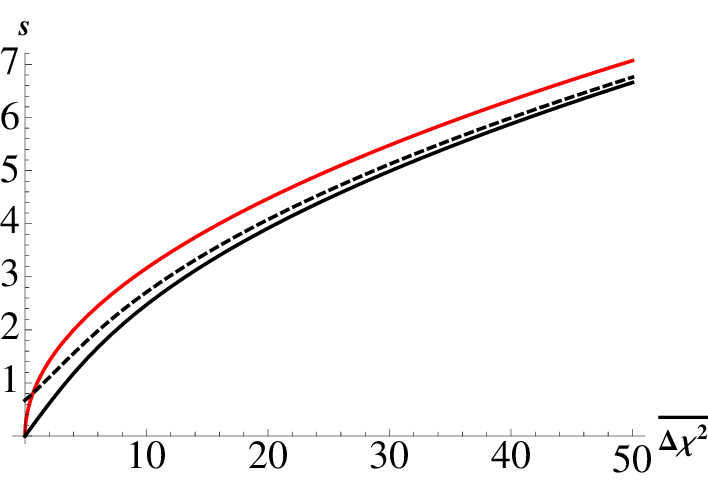}
\caption{For a given $\overline{\Delta\chi^2}$ statistic determined from theoretical spectra, the black curve is the number $s$ of $\sigma$'s of sensitivity of the determination of the mass hierarchy by a median experiment.  A median experiment is one in which $|\Delta\chi^2|$ obtains its median value. For comparison, the dashed curve uses the two-sided definition of $s$ and the red curve is the square root of $\overline{\Delta{\chi^2}}$.}
\label{sfig}
\end{center}
\end{figure}

Using Eq.~(\ref{princ}) one now finds that the number of $\sigma$'s of sensitivity is
\beq
s(\overline{\Delta\chi^2})=\sqrt{2}\ {\rm{erf}}^{-1}\left(\frac{1-e^{-\overline{\Delta\chi^2}/2}}{1+e^{-\overline{\Delta\chi^2}/2}}\right). \label{s}
\eeq
This function is plotted in Fig.~\ref{sfig}.  For example, if $\overline{\Delta\chi^2}=9$ then a median experiment determines the hierarchy with  a sensitivity of 2.3$\sigma$ instead of the $3\sigma$ which might be expected.  Had we insteaded opted for the double-sided Gaussian convention for $s$, we would have instead found 2.5$\sigma$.

A general Bayesian prior of $b$ and $1-b$ for the normal and inverted hierarchies leads to a sensitivity
\beq
s(\overline{\Delta\chi^2})=\sqrt{2}\ {\rm{erf}}^{-1}\left(\frac{1+\left(1-\frac{1}{b}\right)e^{-\overline{\Delta\chi^2}/2}}{1+\left(\frac{1}{b}-1\right)e^{-\overline{\Delta\chi^2}/2}}\right). 
\eeq

\vspace{.2cm}

\noindent
{\it{Third, what is the probability $p(s)$ that the hierarchy will be determined with a sensitivity of at least $s\sigma$?}}

\begin{figure} 
\begin{center}
\includegraphics[width=2.8in,height=1.3in]{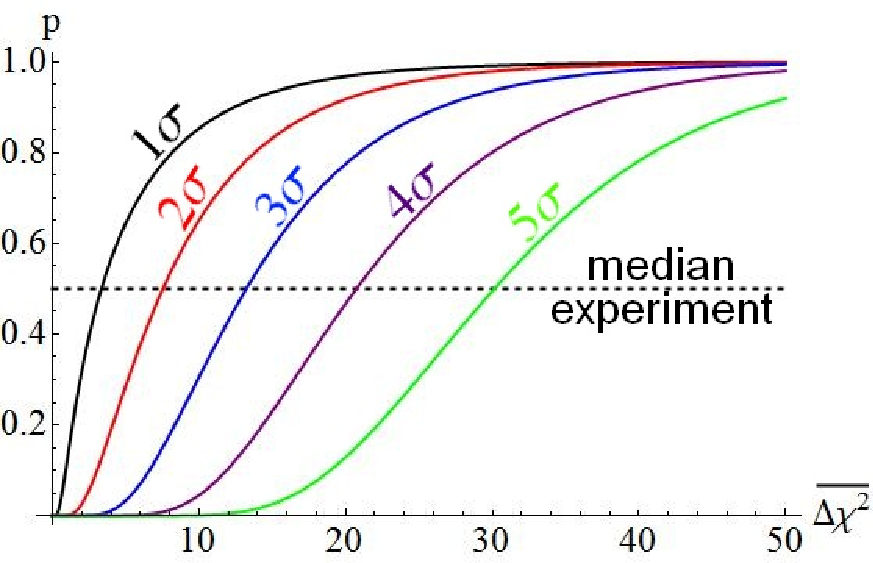}
\caption{The black, red, blue, purple and green curves are the probability of a hierarchy determination with $1\sigma$, $2\sigma$, $3\sigma$, $4\sigma$ and  $5\sigma$ of sensitivity as a function of $\overline{\Delta\chi^2}$.  The dashed line represents a median experiment, and its intersections with the curves yield the same information as Fig.~\ref{sfig}.}
\label{pfig}
\end{center}
\end{figure}

Note first that for a general experimental outcome $\Delta\chi^2$, the probability of success 
\bea
p_v&=&\frac{L_+}{L_++L_-}\nonumber\\&=&\frac{e^{-(\overline{\Delta\chi^2}-\Delta\chi^2)^2/8\overline{\Delta\chi^2}}}{e^{-(\overline{\Delta\chi^2}-\Delta\chi^2)^2/8\overline{\Delta\chi^2}}+
e^{-(\overline{\Delta\chi^2}+\Delta\chi^2)^2/8\overline{\Delta\chi^2}}}\nonumber\\&=&\frac{1}{1+e^{-\Delta\chi^2/2}} 
\eea
is independent of $\overline{\Delta\chi^2}$.  Using this fact, an argument similar to those above leads to
\beq
p(s)=\frac{1}{2}\left(1+{\mathrm{erf}}\left(\frac{\overline{\Delta\chi^2}-4{\mathrm{arctanh}}\left({\mathrm{erf}}\left(\frac{s}{\sqrt{2}}\right)\right)}{\sqrt{8\overline{\Delta\chi^2}}}\right) \right) .
\eeq
This function is plotted in Fig.~\ref{pfig}.

\vspace{.4cm}

\noindent
{\bf{Parallel nuisance parameters}}

In reality there is no single experimental result $\byn$ or $\byi$ which is predicted by a given hierarchy.  The results also depend on a number of nuisance parameters, such as the neutrino mass matrix parameters and the flux normalization of the source.  We will assemble these nuisance parameters into a vector $\bx=\{x_i\}$.

If the final data consists of $N$ numbers, such as the number of events in $N$ energy bins, and if there are $K$ nuisance parameters, then each hierarchy corresponds not to a point but to a $K$-dimensional subset of the $N$-dimensional vector space in which $\by$ is valued.  The nuisance parameters $x_i$ are coordinates on these subsets.  If the standard deviations $\sigma_i$ vary sufficiently slowly, then  the inverse covariance matrix defines a metric on this space.  Recall that, in the case of the normal (inverted) hierarchy, the nuisance parameters $\bx$ are chosen to minimize $\chi^2_N$ $(\chi^2_I)$.  Geometrically, this minimization corresponds to choosing the point in each subset which is closest to $\by$, the coordinates of the point are the nuisance parameters which minimize the corresponding $\chi^2$ statistic.

In this framework, it is easy to combine data from multiple experiments.  They can simply be added to $\by$ as new components.  For example, one can combine a forecast spectrum of Daya Bay II with a value of the nuisance parameter $\theta_{13}$ determined at Daya Bay and RENO by letting the first $N-2$ components of $\by$ correspond to the $\overline{\nu_e}$ spectrum at Daya Bay II and the next two to the relative survival probabilities observed at Daya Bay and RENO.  The single nuisance parameter $\theta_{13}$ yields a curve in the $N$-dimensional space of observations for each hierarchy.  The curve is parameterized by $\theta_{13}$.  The last two coordinates of this curve are simply the relative survival probabilities expected at Daya Bay and RENO as a function of the parameter $\theta_{13}$.  The $\chi^2$ to be minimized is the distance in the full $N$ dimensional space, so it automatically combines determinations of $\theta_{13}$ at RENO, Daya Bay and Daya Bay II without the need for any penalty terms.

Now let us make two approximations.  First, we approximate $\byn$ and $\byi$ to be linear (or affine) functions of the nuisance parameters $\bx$, so that the subspaces corresponding to theoretical predictions are hyperplanes.  The resulting models are called linear regression models.  Model selection in one dimensional non-nested linear regression models was first studied in Ref.~\cite{hotelling1940}.   Ref.~\cite{cox} presented a statistic, generalizing $\Delta\chi^2$, which is Gaussian distributed and distinguishes the models.   The properties of this statistic, in the case of linear regression models, were determined in Ref.~\cite{pesaran1974}.  

One may object that the spectra are not indeed linear functions of the neutrino mass matrix.  However the essential point is that they be approximately linear in a regime whose size is the precision to which an experiment can determine the nuisance parameters.  This is a much easier criterion.  Later we will compare our analytical results to simulations in which no such approximation is made, and we will see that the resulting error is small.

For now we will make the further approximation that one obtains the same value of $\Delta\chi^2$ for any value of the nuisance parameters chosen for the normal hierarchy if the nuisance parameters for the inverse hierarchy are chosen so as to minimize $\Delta\chi^2_I$.  In other words, $\overline{\Delta\chi^2}$, is independent of the choice of the nuisance parameters so long as each $\chi^2$ is properly minimized.  Geometrically this means that the hyperplanes corresponding to the theoretical values $\byn$ and $\byi$ are parallel.  

Again assume that the normal hierarchy is correct.  If ${\mathbf{x}}^{\mathrm{T}}$ is the true value of the nuisance parameters, then the theoretical values of the observables $\byn$ will be linear functions $y_i^{\mathrm{N}}$ of ${\mathbf{x}}^{\mathrm{T}}$.
 $\chi^2_N$ $(\chi^2_I)$ is just the minimum distance squared from the observations $y_i=y_i^{\mathrm{N}}+\sigma_i g_i$ to the hyperplane corresponding to the normal (inverted) hierarchy.  The statistical fluctuation vector ${\mathbf{g}}=\sigma_i g_i$ can be decomposed into a two vectors, ${\mathbf{g^\perp}}$ and ${\mathbf{g^\parallel}}$ such that ${\mathbf{g^\perp}}$ is perpendicular to the hyperplanes and ${\mathbf{g^\parallel}}$ is parallel.

To determine $\chi^2_N$ or $\chi^2_I$, one must choose the nuisance parameters $\bx$ at which it is minimized.   $\chi^2$ will be minimized for the choice of nuisance parameters ${\mathbf{x}}^{\mathrm{T}}+{\mathbf{g^\parallel}}$.  In other words, the parallel part of ${\mathbf{g}}$ yields the statistical error in the determination of the nuisance parameters.  We have assumed that this error is the same for both hierarchies.  For this choice of nuisance parameters, the theoretical predictions for $y_i$ are $y_i^{\mathrm{N}}+\sigma_i g^\parallel_i$ and $y_i^{\mathrm{I}}+\sigma_i g^\parallel_i$ in the cases of the two hierarchies.

\begin{figure} 
\begin{center}
\includegraphics[width=2.8in,height=2.0in]{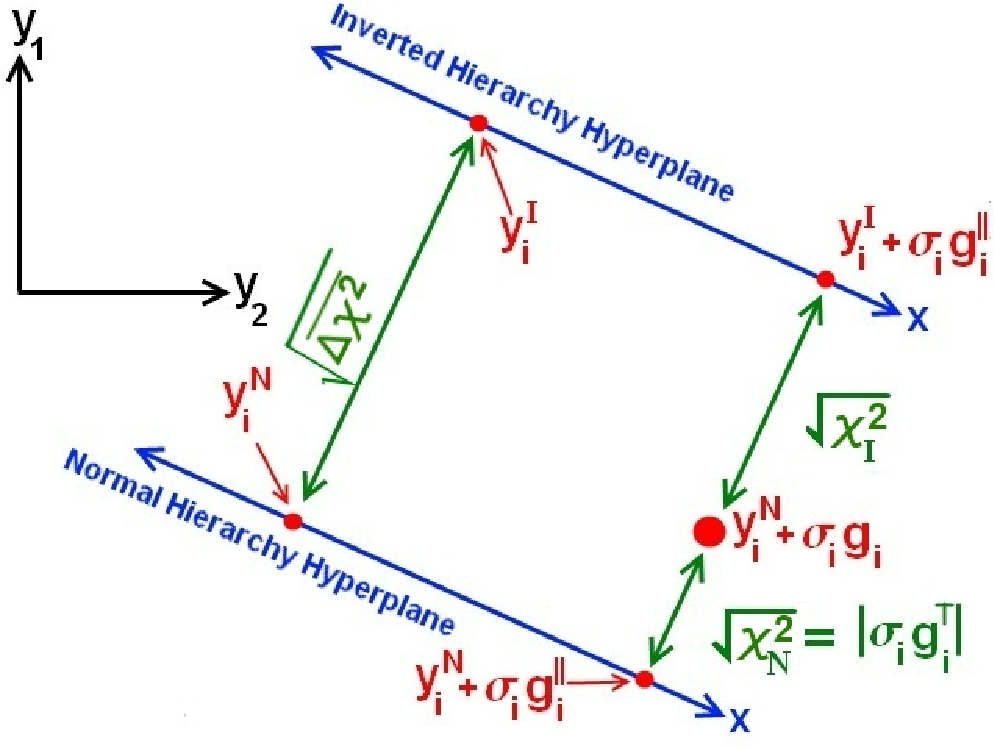}
\caption{In this figure the hierarchy is normal and $\overline{\Delta\chi^2}$ is independent of the nuisance parameters.  The two parallel lines are the expected measurements corresponding to various values of the nuisance parameters for the two hierarchies.  As a result of statistical fluctuations $y_i^{\mathrm{N}}+\sigma_i g_i$ is measured instead of the theoretical value $y_i^{\mathrm{N}}$.  The parallel part of $\mathbf{g}$ determines the effect of this fluctuation on the best fit nuisance parameters and the perpendicular part its effect on $\Delta\chi^2$.}
\label{NOvA}
\end{center}
\end{figure}

Now we are ready to calculate
\bea
\Delta\chi^2&=&\chi^2_I-\chi^2_N \label{xincontb}\\
&=&\sum_i\frac{(y_i-y_i^{\mathrm{I}}-\sigma_i g^\parallel_i)^2}{\sigma_i^2}-\sum_i\frac{(y_i-y_i^{\mathrm{N}}-\sigma_i g^\parallel_i)^2}{\sigma_i^2}\nonumber\\
&=&\sum_i\frac{(y_i^{\mathrm{N}}+\sigma_i g_i^\perp-y_i^{\mathrm{I}})^2-(y_i^{\mathrm{N}}+\sigma_i g_i^\perp-y_i^{\mathrm{N}})^2}{\sigma_i^2}\nonumber\\
&=&\sum_i\frac{(y_i^{\mathrm{N}}-y_i^{\mathrm{I}})^2}{\sigma_i^2}+\sum_i\frac{2(y_i^{\mathrm{N}}-y_i^{\mathrm{I}})}{\sigma_i}g_i^\perp.\nonumber\\
&=&\sum_i\frac{(y_i^{\mathrm{N}}-y_i^{\mathrm{I}})^2}{\sigma_i^2}+\sum_i\frac{2(y_i^{\mathrm{N}}-y_i^{\mathrm{I}})}{\sigma_i}g_i.\nonumber
\eea 
In the last step we used the identity
\beq
\sum_i\frac{2(y_i^{\mathrm{N}}-y_i^{\mathrm{I}})}{\sigma_i}g_i^\parallel=0
\eeq
which follows from the fact that, using the metric $1/\sigma_i^2$, the vector $(y_i^{\mathrm{N}}-y_i^{\mathrm{I}})$ is perpendicular to the hyperplanes and so to $\sigma_i g_i^\parallel$.

Just as in Eq.~(\ref{xincont}), Eq.~(\ref{xincontb}) describes a normal distribution centered at $\overline{\Delta\chi^2}$ and with standard deviation $2\sqrt{\overline{\Delta\chi^2}}$.  As a result, Eqs.~(\ref{suc}) and (\ref{s}) for the probability of success and number of $\sigma$'s of sensitivity in the median experiment remain correct.

\vspace{.4cm}

\noindent
{\bf{General nuisance parameters}}

Of course, $\overline{\Delta\chi^2}$ does depend on the nuisance parameters, and so the hyperplanes corresponding to the theoretical data are not parallel and the above results are only approximate.  This fact was first noted in Ref.~\cite{cox}, where it was concluded that as a result $\Delta\chi^2$ is not normally distributed.  Its distribution leptokurtic.   

This observation can be intuitively understood as follows.  Imagine that $\overline{\Delta\chi^2}$ depends so strongly upon the nuisance parameters that a 1$\sigma$ change in the nuisance parameters can reduce the sensitivity to the hierarchy by several $\sigma$'s.  As a result, most of the experiments in which the hierarchy determination is incorrect will be those in which the nuisance parameter is such that $\overline{\Delta\chi^2}$ is much smaller.  Thus the tails of the distribution of $\Delta\chi^2$ will grow as a result of those simulations in which the nuisance parameters take a nonstandard value.  Clearly, this effect is only present in simulations in which the nuisance parameters are allowed to vary, and so simulations that fix the nuisance parameters will yield values of $\Delta\chi^2$ which, upon using Eq.~(\ref{s}), overestimate the sensitivity to the hierarchy.  

In Ref.~\cite{cox} the author proposed a new statistic which does follow a Gaussian distribution even in this more general setting.  However, in the case of the hierarchy determinations planned in the near future, the angle between the hypersurfaces is actually quite small.   This is reflected in the observation \cite{noisim2} that even a 1$\sigma$ variation in $\theta_{13}$ only leads to about a one third of a $\sigma$ variation in the confidence.  Therefore the approximate treatment of the $\Delta\chi^2$ statistic above is quite precise.

\begin{figure} 
\begin{center}
\includegraphics[width=2.8in,height=1.2in]{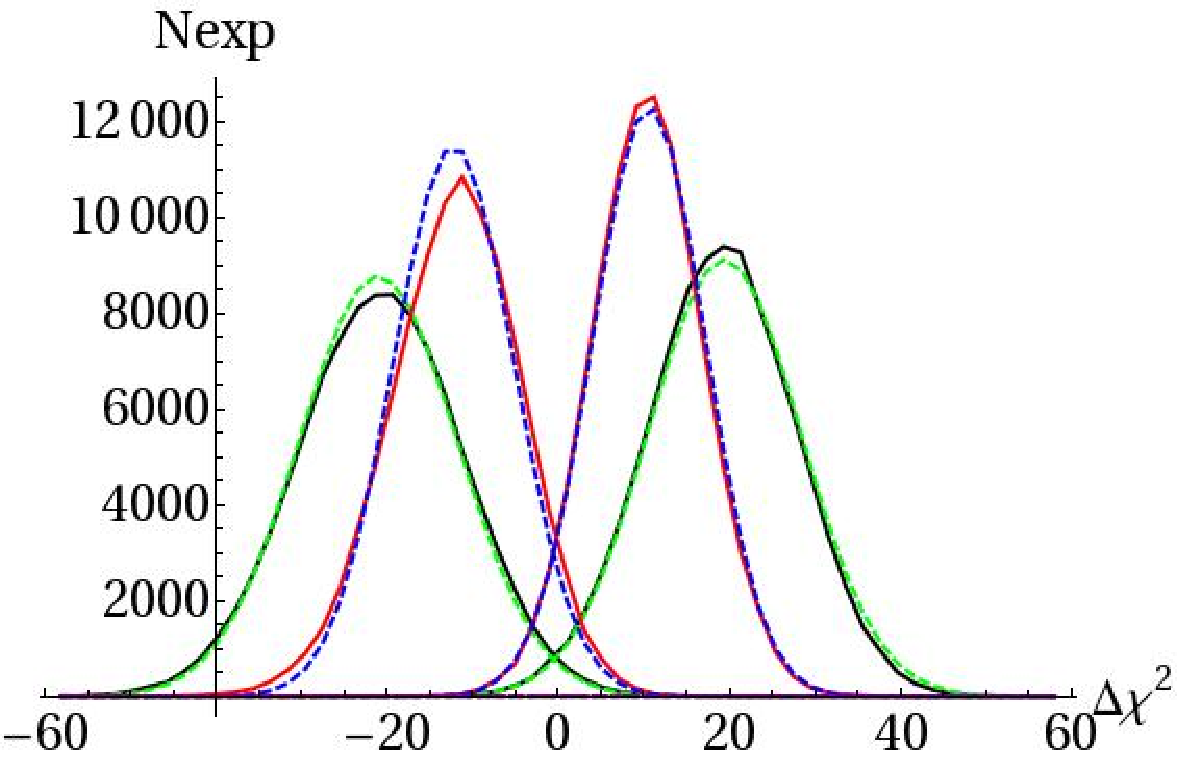}
\caption{The distribution of $\Delta\chi^2$ in 50,000 experiments with each hierarchy is shown, combining the data with MINOS' 4\% determination of the atmospheric mass splitting (red curve) and with an optimistic 1\% determination at NO$\nu$A (black curve).  The dashed curves are the corresponding Gaussian distributions centered at $\overline{\Delta\chi^2}$ with width $2\sqrt{\overline{\Delta\chi^2}}$.}
\label{NOvAfig}
\end{center}
\end{figure}

To illustrate this point, in Fig.~\ref{NOvAfig} we present the distribution of the $\Delta\chi^2$ statistic in simulations which combine the $\overline{\nu}_e$ spectrum measured at Daya Bay II with MINOS' 4\% determination of the atmospheric mass difference \cite{parke2005} and also with an optimistic 1\% forecast determination at an upgraded NO$\nu$A .    All of the nuisance parameters are fixed except for $\mn32$, which is chosen to minimize $\chi^2_I$ and $\chi^2_N$ as described above.   Following \cite{yifangmar} we have considered 6 years of exposure at a 20 kton detector for Daya Bay II which detects $\overline{\nu}_e$ via inverse $\beta$ decay on the 10\% of its mass consisting of free protons.  The baselines and reactor fluxes are identical to Ref.~\cite{yifangmar}.  The leptonic CP-violating angle $\delta$ is set to $\pi/2$.   

We find that the distribution of $\Delta\chi^2$ is indeed well approximated by a Gaussian distribution centered at $\overline{\Delta\chi^2}$ with standard deviation $2\sqrt{\overline{\Delta\chi^2}}$.    $\overline{\Delta\chi^2}\sim 11\ (20)$ for Daya Bay II with MINOS (NO$\nu$A) yielding $2.6\sigma$ ($3.9\sigma$) of sensitivity at the median experiment, with a rate of successfully determining the hierarchy of 94.6\% (98.5\%) in good agreement with Eq.~(\ref{suc}).   

\begin{figure} 
\begin{center}
\includegraphics[width=2.8in,height=1.2in]{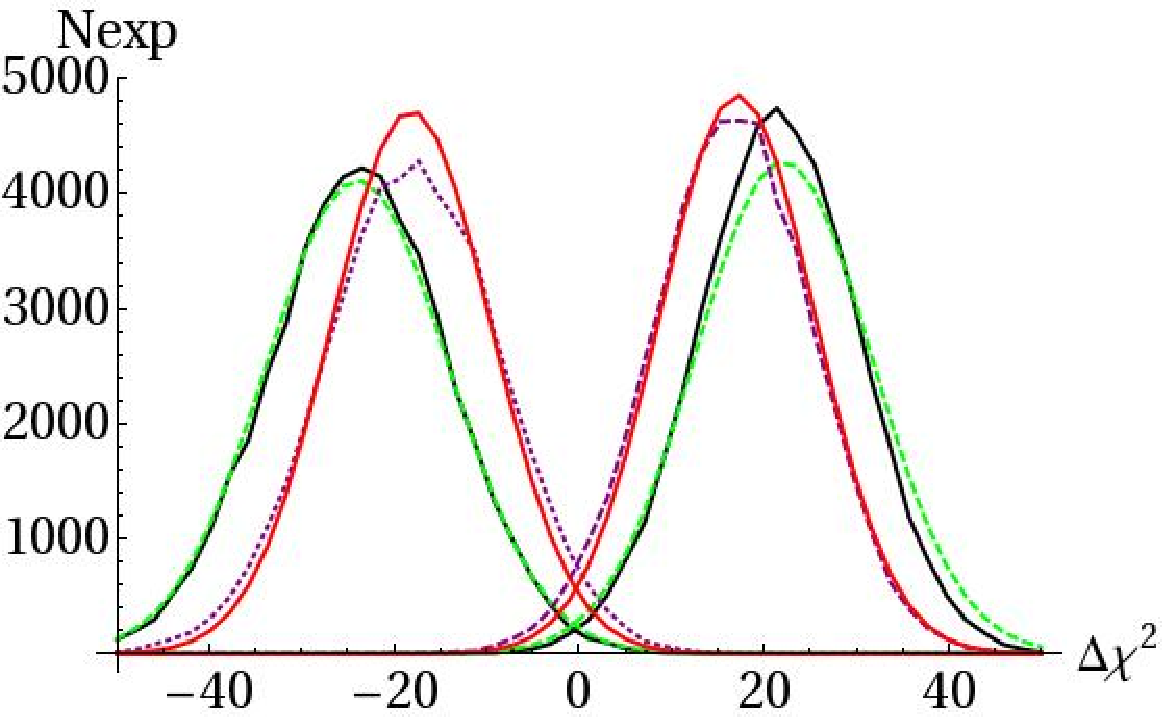}
\caption{As in Fig.~\ref{NOvAfig}, but using only a 1\% determination of the atmospheric mass splitting.  The simulations reported in the red and black curves use $\delta=0$ and $\pi$ respectively, although the fitting is always performed assuming $\delta=\pi/2$.  As can be seen, if $\delta=\pi$, the hierarchy determination will be more reliable \cite{noisim2,yifangmar}.}
\label{cp}
\end{center}
\end{figure}

In Fig.~\ref{cp} we present the distribution of $\Delta\chi^2$ in simulations in which $\delta=0$ and $\pi$, although we always fit to a $\delta=\pi/2$ theoretical mode as the appearance mode at T2K and NO$\nu$A cannot distinguish $0$ and $\pi$ \cite{noisim2,parke2013}.  At $\delta=0$ ($\pi$) we find $\overline{\Delta\chi^2}=17$ (22) yielding $3.5\sigma$ (4.2$\sigma$) of sensitivity, confirming the expectations of Ref.~\cite{minakata}.  Despite the fact that the model used for fitting differs from that used to generate the data, the distribution of $\Delta\chi^2$  described in this paper approximates the simulated data well.

\vspace{.4cm}

\noindent
{\bf{Frequentist confidence}}

A frequentist notion of confidence can be made well defined even in this context \cite{yosh,pingu}.  Imagine that an experiment measures $\Delta\chi^2$.  This differs from the expected $\overline{\Delta\chi^2}$ for the normal (inverted) hierarchy by $|{\overline{\Delta\chi^2}}\mp\Delta\chi^2|$ which corresponds to a frequentist incompatibility of
\beq
\frac{|{\overline{\Delta\chi^2}}\mp\Delta\chi^2|}{2\sqrt{\overline{\Delta\chi^2}}}
\eeq
$\sigma$'s.

In particular, in the case of the median experiment with the true hierarchy, $\Delta\chi^2={\overline{\Delta\chi^2}}$.  Therefore the inverted hierarchy is excluded at a confidence of $\sqrt{\Delta\chi^2}$\ $\sigma$'s.  In this sense it might be tempting to ignore the results of this paper and to identify the frequentist incompatibility $\sqrt{\Delta\chi^2}$ with the confidence in the hierarchy determination expected in a median experiment.

While such a definition of confidence is well-defined, it has a very unattractive feature.  Consider an experiment with an expected ${\overline{\Delta\chi^2}}=16$.  The general arguments in this note imply that if the hierarchy is normal (inverted) then $\Delta\chi^2$ will follow a Gaussian distribution centered on $16\ (-16)$ with a width of $\sigma=8$.  In the frequentist sense, the median experiment will yield $|\Delta\chi^2|=16$ and so is incompatible with the false hierarchy with $4\sigma$ of confidence while the 98th percentile experiment will yield $\Delta\chi^2=0$ and so is incompatible with the false hierarchy with $2\sigma$ of confidence.  An identification of the sensitivity to the hierarchy with the frequentist incompatibility would therefore imply that even the 98th percentile of experimental outcomes will yield a $2\sigma$ sensitivity to the hierarchy.

Now consider the somewhat unlikely case in which due to statistical fluctuations, the results of this experiment are indeed in the 98th percentile, so that $\Delta\chi^2=0$.  Now the experimentalist will be asked to provide the hierarchy with $2\sigma$ of confidence.  Of course he cannot, the experiment has not yielded any preference for either hierarchy, even at the $2\sigma$ level that was promised for a 98th percentile experiment when the funding was requested.  In this sense, the identification of the frequentist incompatibility with the confidence in the hierarchy determination is misleading: the confidence can be nonzero even when no information is obtained.

The basic problem with the application of the frequentist notion of confidence in this example is that both hierarchies have been ruled out with equal confidence.  Ruling out both hierarchies can be useful when searching for new physics, testing assumptions regarding backgrounds, etc.  Although in that case one would use a $\chi^2$ test and not a $\Delta\chi^2$ test, as the latter is insensitive to effects that affect both hierarchies similarly.  However, for the purpose of determining which hierarchy is manifested in nature it is reasonable to assume that one of the hierarchies is indeed correct.  In this case one is led to the Bayesian constructions described in this note.

\section* {Acknowledgement}

\noindent
We have benefited beyond measure from correspondence with Xin Qian, Yoshitaro Takaesu and Tyce DeYoung.  We thank Amir Nawaz Khan for identifying two typos in an earlier version of this paper.  JE is supported by the Chinese Academy of Sciences Fellowship for Young International Scientists grant number
2010Y2JA01. EC and XZ are supported in part by the NSF of
China.   


\end{document}